# Band splitting with vanishing spin polarizations in noncentrosymmetric crystals

Kai Liu[1,2], Wei Luo[1,2], Junyi Ji[1,2], Paolo Barone[3], Silvia Picozzi[3]*, Hongjun Xiang[1,2]*

[1]*Key Laboratory of Computational Physical Sciences (Ministry of Education), State Key Laboratory of Surface Physics, and Department of Physics, Fudan University, Shanghai 200433, P. R. China*

[2]*Collaborative Innovation Center of Advanced Microstructures, Nanjing 210093, P. R. China*

[3]*Consiglio Nazionale delle Ricerche CNR-SPIN Via dei Vestini 31, Chieti 66100, Italy*

E-mail: silvia.picozzi@spin.cnr.it, hxiang@fudan.edu.cn

**Abstract:** The Dresselhaus and Rashba effects are well-known phenomena in solid-state physics, in which spin–orbit coupling (SOC) splits spin-up and spin-down energy bands of nonmagnetic non-centrosymmetric crystals. Here, we discover a new phenomenon, dubbed as band splitting with vanishing spin polarizations (BSVSP), in which, as usual, SOC splits the energy bands in nonmagnetic non-centrosymmetric systems; surprisingly, however, both split bands show no net spin polarization along certain high-symmetry lines in the Brillouin zone. In order to rationalize this phenomenon, we propose a new classification of point groups into pseudo-polar and non-pseudo-polar groups. By means of first-principles simulations, we demonstrate that BSVSP can take place in both symmorphic (e.g., bulk GaAs) and non-symmorphic systems (e.g., two dimensional ferroelectric SnTe). Furthermore, we propose a novel linear magnetoelectric coupling in reciprocal space, which could be employed to tune the spin polarization with an external electric field. The BSVSP effect and its manipulation could therefore pave a new way to novel spintronic devices.

The study of a relativistic interaction, i.e., the spin-orbit coupling (SOC), has been one of the central themes in the context of spintronics, branch of electronics aiming at utilizing the electron's spin degree of freedom for device applications.[1] Many intriguing SOC-related phenomena were observed, such as spin relaxation[2,3], optical spin orientation[4], spin Hall effects[5-8], persistent spin structures[9-11], hidden spin polarization in centrosymmetric systems[12], and the spin galvanic effects[13,14]. In particular, Dresselhaus[15] and Rashba[16-18] demonstrated that SOC splits spin-up and spin-down bands in nonmagnetic systems lacking inversion symmetry, inducing an effective magnetic field that depends on the crystal momentum $k$. Although the Dresselhaus and Rashba effects lead to different spin polarizations of the energy bands, they result in a similar band dispersion: the original spin degenerate bands split so that spin-up and spin-down bands shift towards opposite directions in $k$ space, as shown in Fig. 1. Even though the Rashba effect is generally associated to heterostructures and surfaces, as due to 2D structural inversion asymmetry, a giant Rashba effect has been recently observed in bulk polar/ferroelectric systems[19-23]. Sometimes, Dresselhaus and Rashba terms coexist[24], giving rise to interesting physics and novel applications. For example, if the magnitudes of Dresselhaus and Rashba terms are equal, the spin–orbit field is unidirectional, resulting in a momentum-independent spin configuration (i.e., the so-called "persistent spin texture"), which could support an extraordinarily long spin lifetime of carriers[9]. To further develop the field of spin–orbitronics (a new branch of spintronics)[25], it is of great interest to explore new SOC-related phenomena beyond the well-known Dresselhaus-Rashba effects. Indeed, a first contribution in this direction has been recently provided by a careful analysis of bulk and site symmetries of crystalline materials, that allowed for a unified description of spin-splitting effects and, more importantly, to unveil hidden spin-polarization phenomena in centrosymmetric systems[12], recently observed in several layered compounds[26-32].

In this work, we discover a novel phenomenon, i.e., band splitting with vanishing spin polarizations (BSVSP) in acentric nonmagnetic materials, as shown in Fig. 1. The group theory analysis indicates that the band splitting arises from the breaking of inversion symmetry and the vanishing spin polarization (i.e., the expectation value of spin evaluated over each Bloch wavefunction) is due to the presence of additional symmetries. We demonstrate in details why BSVSP is possible and how to achieve it in two prototypical examples, namely bulk GaAs and two-dimensional ferroelectric SnTe, complementing the symmetry analysis with first-principles calculations [see section I of supplementary material (SM)].

**Group theory analysis on BSVSP.** To facilitate our discussion, we first classify the 32 crystallographic point groups into pseudo-polar and non-pseudo-polar point groups (Fig. 2), where a pseudo-vector (such as spin, magnetization, angular moment, magnetic field) is allowed or forbidden, respectively. This classification (proposed in this work for the first time to our best knowledge) is in the same spirit as the classification of the point groups into polar and non-polar point groups. As we will show below, it will be very useful for checking the existence of bulk and local spin polarizations. We note that spin vectors $S_x$, $S_y$ and $S_z$ will not belong to the identical representation for a non-pseudo-polar point group.

The essence of BSVSP manifests in a non-degenerate band at a given crystal momentum $k$ lacking net spin polarizations in an inversion-asymmetry crystal, as shown in Fig. 1. As such, the BSVSP effect cannot be realized in nonmagnetic centrosymmetric crystals, where Kramers theorem implies that all bands are at least two-fold degenerate (spin degeneracy). Note that BSVSP does not mean the Bloch states are spinless since the expectation value of $S^2$ is not zero. In order to realize BSVSP at a given momentum $k$, two conditions should be simultaneously satisfied: (1) The eigenstate - including SOC - should be non-degenerate, suggesting that the little space group associated with $k$ should possess at least one 1D double-valued IR; (2) The little point group associated with $k$ should be a non-pseudo-polar point group. Detailed analysis based on group theory (see section III of SM) suggests that BSVSP can not only occur in systems with symmorphic space groups, but also in systems with non-symmorphic space groups. In Table I, we list all the possible 4 symmorphic space groups which might display BSVSP. The corresponding high-symmetry lines where BSVSP takes place are also given. For a given non-symmorphic space group, one first finds out the possible $k$-points on the Brillouin zone boundary where the corresponding little point group has at least one 1D double-valued IRs. Then, among these $k$-points, one finds out the subset of the $k$-points whose corresponding little point group is non-pseudo-polar. To detect BSVSP in experiment, one can use, for example, spin-polarized angle-resolved photoelectron spectroscopy[26] to measure the spin polarizations of the bands along a special high symmetry line.

Although the net spin polarizations are zero for both split bands in the BSVSP effect, a local spin polarization might exist, in loose analogy with the hidden spin-polarization effect. By projecting the spin polarization on atomic sites in real space, it is possible to evaluate the contribution of magnetization of each atom to a chosen band. The magnetization of each atom must satisfy the symmetry of the intersection of the site symmetry group and little point group of the considered momentum $k$. If the resulting symmetry group is non-pseudo-polar, the local spin polarizations also vanish on the chosen atom. Otherwise, local spin polarizations could survive. We find that both situations can occur [GaAs (see below) and SnI4 (see section V of SM) are representative of the two cases, respectively].

To prove the existence of BSVSP, we calculated from first principles the band structures and related spin polarizations of bulk GaAs and 2D SnTe, representative of symmorphic and non-symmorphic case, respectively.

**BSVSP in zinc-blend GaAs.** Bulk GaAs shows a zinc-blende structure with $\bar{F}43m$ space group whose corresponding point group is $T_\text{d}$. There is no inversion symmetry in GaAs, but since the crystallographic point group is non-polar, it may exhibit Dresselhaus effect (D-1, according to the classification given in Ref. 12). In Fig. 3(a), we show the calculated SOC band structures and the $x$ components of spin polarizations dependence on $k$ in the vicinity of the Γ point for GaAs. The $y$ and $z$ components of spin polarizations have the same behavior. It is well-known that the top valence bands are four-fold degenerate at Γ. Along symmetry line Λ [Γ → L(0.5,0.5,0.5)], the top valence bands split into two non-degenerate bands (labeled by $\widetilde{\Lambda}_4$ and $\widetilde{\Lambda}_5$, respectively) and a two-fold degenerate band (labeled by $\widetilde{\Lambda}_6$). Note that the two-fold degenerate bands are composed by spin-up and spin-down components, resulting in a zero net spin polarization. Interestingly and surprisingly, the top two non-degenerate bands also have vanishing spin polarization, i.e., displaying the BSVSP behavior. Note that all bands are at least two-fold degenerate for isostructural bulk Si with the inversion symmetry. This means that inversion asymmetry leads to the splitting of the top valence band in GaAs. It should be noted that BSVSP takes place only along special high-symmetry lines. In fact, conventional Dresselhaus splitting with spin polarization takes place for other symmetry lines [e.g., symmetry line Γ → A(0.5,0.25,0), as shown in Fig. 3(a)].

The vanishing spin polarizations can be rationalized by considering the band symmetry properties. Along the Λ symmetry line, the little point group is $C_{3v}$. According to our previous analysis, $C_{3v}$ is classified as non-pseudo-polar. Therefore, the $x$, $y$ and $z$ components of spin polarizations must vanish. Here, we illustrate why the spin polarization along the symmetry line Λ should vanish in another clear way. Adopting a xyz coordinate system with the local z-axis along the [111] direction of the cubic cell of bulk GaAs, the point group $C_{3v}$ has six symmetry operations, including a mirror symmetry $M_1$ which consists of reflection about the $y = 0$ plane

$$M_1: (S_x, S_y, S_z) \rightarrow (-S_x, S_y, -S_z),$$

and a three-fold rotation $C_{3z}$ which consists of π/3 rotation around the $z$ axis:

$$C_{3z}: (S_x, S_y, S_z) \rightarrow \left(-\frac{1}{2}S_x - \frac{\sqrt{3}}{2}S_y, \frac{\sqrt{3}}{2}S_x - \frac{1}{2}S_y, S_z\right),$$

where, $S_x$, $S_y$ and $S_z$ are the cartesian components of the spin polarization. The reflection $M_1$ ensures the vanishing $x$ and $z$ components of spin polarizations, while the three-fold screw rotation $C_{3z}$ ensures the vanishing $x$ and $y$ components of spin polarizations. Overall, the spin polarization should be zero.

The top valence bands of zinc-blend semiconductors are well described by the 4×4 Luttinger model expressed in the basis of total angular momentum $j = 3/2$.[33,34] We further adopt this method to prove the unusual band splitting with vanishing spin polarization in GaAs (see section IV of SM). The band structure of the effective Hamiltonian along Λ line are shown in Fig. 3(a). The top valence bands are two non-degenerate bands. The $x$ component of spin polarizations is zero for these non-degenerate bands, as shown in Fig. 3(b). The $y$ and $z$ components have the same behavior. This ***k·p*** analysis confirms our DFT results and group theory analysis. In Dresselhaus/Rashba effect, the spin polarization of the split bands arises as a consequence of the effective k-dependent magnetic field experienced by electrons in the presence of SOC in an acentric environment. When the symmetry requirements unveiled by our group-theory analysis are met, however, the effect of such SOC-induced magnetic fields vanishes, thus giving rise to BSVSP effect. Our further analysis shows indeed that the SOC-induced effective magnetic fields acting on two different $p$ orbitals (e.g., $p_x$ and $p_y$) of the same atom (e.g., As) are equal in strength but opposite in direction for the k-point along the Γ-L direction, resulting in a band splitting and a net vanishing spin polarization (since local magnetic moments for different orbitals of the same atom are opposite) (see Fig.1b and section IV of the SM).

Our calculations show that not only the net spin polarization, but also the local spin polarizations of Ga and As atoms for these two single degenerate bands vanish. This can be understood by symmetry arguments. The site symmetry groups of Ga and As atoms are $T_d$ and the little group of Λ line is $C_{3v}$, which is a subgroup of $T_d$. The magnetization of Ga and As atoms must satisfy the symmetry of $C_{3v}$ which is a non-pseudo-polar point group.

We propose that one can tune spin polarization with strain. For example, an in-plane strain applied to GaAs would lower the crystallographic point group to $D_{2d}$. The little point group along symmetry line Γ→ P (also along [111] direction) is $C_2$ which is pseudo-polar. As a consequence, a net spin polarization is symmetry allowed, as indeed confirmed by our DFT calculations when applying an in-plane compressive strain to cubic GaAs (Fig. S2).

**BSVSP in 2D ferroelectric SnTe.** Recently, 2D ferroelectric SnTe with high Curie temperature has been observed.[35,36] This material comprises two sheets of alternating Sn and Te atoms, where the in-plane ferroelectric polarization – arising from the relative offset of Sn and Te sublattices – breaks the inversion symmetry. Its space group is Pmn$2_1$ with $C_{2v}$ crystallographic point group, where we choose the polarization direction as the $z$ axis, while the direction normal to the 2D thin film is taken as the $y$ axis [see Fig. 4(b)]. In Fig. 4(a) we show the band structure around the X point, where four non-degenerate bands are found along the symmetry lines Σ (X → U) and Δ (X → Γ), merging in two two-fold degenerate bands at the time-reversal invariant point X, labeled as $\widetilde{X}_2$, $\widetilde{X}_5$ and $\widetilde{X}_3$, $\widetilde{X}_4$, respectively. Along the Σ symmetry line, all components of spin polarization vanish, thus realizing the BSVSP effect. On the other hand, the bands along the Δ symmetry line are fully spin polarized along the $y$ axis, as expected for a unidirectional Rashba effect, being the bands spin-polarization perpendicular to both $k$ and ferroelectric polarization. This can be explained by looking at the little groups of the two symmetry lines, following our previous analysis. Along symmetry line Σ, the little point group is $C_{2v}$, which is non-pseudo-polar. Therefore, the net spin polarizations vanish. Since the little point group of Δ symmetry line is the pseudo-polar $C_s$, a non-zero spin polarization is allowed along this line. We further construct wave functions using representation theory to prove the vanishing spin polarizations along the symmetry line Σ (see section VII of SM). A minimal $k·p$ model can be written for the valence top bands around X, which reads as $H = -\Delta\tau_y\sigma_y - \alpha_B(k_z\tau_z\sigma_x - k_x\tau_z\sigma_z) + \beta_R k_x \tau_0 \sigma_y + \alpha_R k_x \tau_y \sigma_0$, where $2\Delta$ is the SOC induced splitting at X, $\alpha_B$ ($\alpha_R$ and $\beta_R$) measure the strength of BSVSP (Rashba-like effect), $\sigma$ and $\tau$ are Pauli matrices representing the spin and pseudospin degrees of freedom, respectively (the latter spanning the two-dimensional single-valued representation $\widetilde{X}_1$, see section VII of SM). The $k·p$ model also predicts the BSVSP effect.

Spin polarizations along the Σ symmetry line projected on Sn and Te atoms are shown in Fig. 4(b). For each atom, only the $x$ component of the spin polarization does not vanish. This is rationalized by considering the site symmetries of Sn and Te atoms. The intersection of site symmetry groups ($C_s$, comprising a $M_{yz}$ mirror operation about the $x = 0$ plane) and little point group of Σ line ($C_{2v}$) is $C_s$ which is pseudo-polar. Therefore, there are spin polarizations on each atom. Only the $x$ components survive as the glide reflection $M_{yz}$ results in the zero $y$ and $z$ components. However, local spin polarizations projected on different Sn and Te atoms have the same values but opposite sign. They cancel each other so the total spin polarizations vanish. Furthermore, all local spin polarizations are reversed in the conjugate split band. This behavior can be regarded as "antiferromagnetism of a Bloch state". The antiferromagnetic order satisfies the little space group symmetry of the high-symmetry line Σ. In the 2D SnTe case, the magnetic fields acting on different atoms of the same kind (e.g., Sn) are opposite in direction for the k-point along the X-U direction, resulting in a band splitting and a net vanishing spin polarization (since local magnetic moments for the same kind atom at different positions are opposite) (see Fig. 1b and section VII of the SM). The presence of a staggered k-dependent magnetic field along the X-U direction clearly appears in the k.p model restricted to the Σ line, which can be recast as $H = -\Delta\tau_y\sigma_y + B_x(k)\sigma_x$, where $B_x(k) = -\tau_z\alpha_B k_z$.

**Tuning of spin polarization with electric field.** Previously, it was proposed that the electric field can switch the electric polarization, thus tune the spin texture[21,37-39]. In the 2D SnTe case, the switching of the in-plane electric polarization by 90° or 180° by electric field can also result in a change of the spin texture. Here, we rather propose a new electric field effect. Since the BSVSP effect is protected by symmetries (e.g., mirror plane), one can use an electric field to break these symmetries and induce a net spin polarization. For example, the SnTe vanishing spin polarizations are protected by the two reflections, $y = 0$ and $x = 0$ planes. Applying an external electric field along the $y$ axis (normal to the 2D thin film), which can break the glide reflection $y = 0$ plane and reduce the symmetry of 2D SnTe thin film to the pseudo-polar $C_s$ symmetry, results in the emergence of the $x$ component of spin polarizations. As shown in Fig. 5(a), the dependence of the $x$ components on $k$ by applying a 0.1 V/ Å external electric field indeed confirms our analysis. By projecting spin polarizations on atoms, one can observe that the $x$ components of spin polarizations with different orientations do not cancel each other any more, as shown in Fig. 5(b). This behavior can be regarded as "ferrimagnetism of the Bloch state" and ensures the emergence of spin polarizations. The magnitude of spin polarization at a $k$ point is linearly proportional to the external electric field, as shown in Fig. 5(c). In particular, an opposite external electric field can lead to a flop of spin polarizations, as shown in Fig. S5. This can be understood since these two states are related by the glide reflection - $y = 0$ plane. Our k.p model analysis shows that this electric field effect arises from the fact that the electric field causes an energy splitting of the orbitals located at different positions along the out-of-plane direction (see section VIII of the SM). In some sense, this effect can be considered as a new type of linear magnetoelectric coupling[40], namely linear magnetoelectric coupling in reciprocal space. We recall that in the conventional linear magnetoelectric coupling, the electric field induces a change of magnetization in the whole system. On the contrary, in the linear magnetoelectric coupling in reciprocal space hereby proposed, the electric field induces a change of spin polarization for some particular Bloch eigenstates.

The BSVSP effect might lead to some new spintronic applications. By applying strain (e.g., via a piezoelectric material) or electric field to the system displaying the BSVSP effect, the spin polarization can be adjusted, thus the spin transport behavior can be tuned, which might lead to novel spintronic devices (such as spin-FET[41-44]).

In summary, we put forward a new phenomenon (dubbed as BSVSP) in nonmagnetic inversion-asymmetric systems: band splitting induced by inversion-symmetry breaking, while the net spin polarization is zero. This phenomenon is rationalized in terms of protection by non-pseudo-polar symmetries. The BSVSP behavior can occur not only in symmorphic but also in non-symmorphic systems, as shown by our density functional theory calculations demonstrating that BSVSP takes place in bulk GaAs and 2D ferroelectric SnTe. BSVSP is not only interesting from the physics point of view, but also promising for novel applications. For instance, we propose a novel linear magnetoelectric coupling in reciprocal space, which can be adopted to tune the spin polarizations of Bloch states by means of an external electric field.

Table I. All possible symmorphic space groups that a crystal displaying BSVSP might belongs to. The corresponding crystal point groups and high symmetry lines at which BSVSP occurs are also listed. Here, "$u$" characterizes the position of the k-point in the high symmetry line.

| Crystal point groups | Symmorphic space groups | High symmetry lines |
|---|---|---|
| $3m$ ($C_{3v}$) | $P3m1$, $P31m$, $R3m$ | $(0,0,u)$ |
| $\bar{6}m2$ ($D_{3h}$) | $P6mm$ | $(1/3,1/3,u)$ |
| $6mm$ ($C_{6v}$) | $P\bar{6}m2$, $P\bar{6}2m$ | $(0,0,u)$, $(1/3,1/3,u)$ |
| $\bar{4}3m$ ($T_d$) | $P\bar{4}3m$, $F\bar{4}3m$, $I\bar{4}3m$ | $(u,u,u)$ |

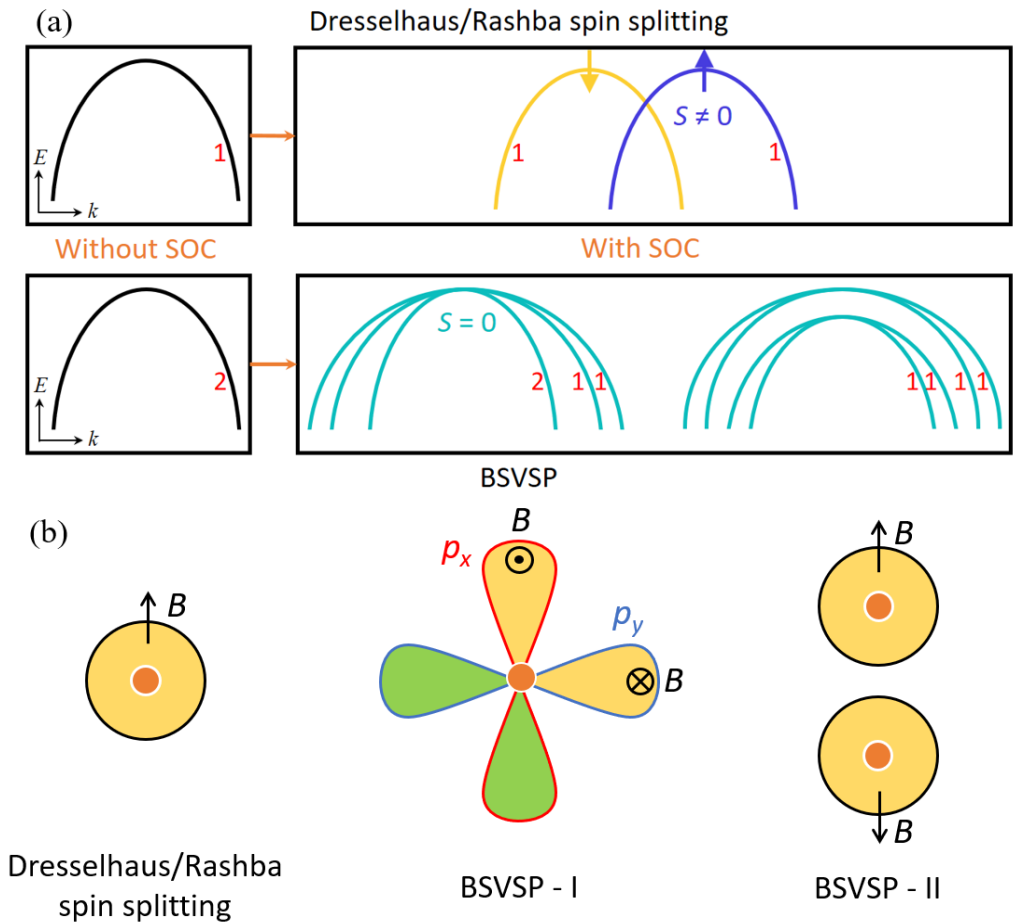

Figure 1. (a) Schematic depiction of Dresselhaus/Rashba spin splitting and BSVSP. "1" and "2" means single and double degenerate, respectively. In the Dresselhaus/Rashba spin splitting case, a non-degenerate band (without considering spin) splits into two

bands with opposite nonzero spin polarizations. In the case of BSVSP, a double degenerate band (without considering spin) splits, resulting in some non-degenerate bands without net spin polarizations. (b) Schematic illustration of the difference between Dresselhaus/Rashba spin splitting and BSVSP. In the case of Dresselhaus/Rashba spin splitting, a k-dependent magnetic field induces the band splitting. In BSVSP–I (II), the magnetic fields acting on two different $p$ orbitals of the same atom (acting on different atoms of the same kind) are equal in strength but opposite, resulting in a band splitting with vanishing spin polarization.

| Pseudo-polar | | | Non-pseudo-polar | | | |
|---|---|---|---|---|---|---|
| $C_1$ | $C_2$ | $C_3$ | $C_{2v}$ | $C_{3v}$ | $C_{4v}$ | $C_{6v}$ |
| $C_4$ | $C_6$ | $C_s$ | $D_2$ | $D_3$ | $D_4$ | $D_6$ |
| $C_{3h}$ | $S_4$ | $C_{2h}$ | $D_{3h}$ | $D_{2d}$ | $T$ | $T_d$ |
| $C_{4h}$ | $C_{6h}$ | $C_i$ | $O$ | $D_{2h}$ | $D_{4h}$ | $D_{6h}$ |
| $S_6$ | | | $D_{3d}$ | $T_h$ | $O_h$ | |

Figure 2. Classification of 32 crystal point groups into pseudo-polar and non-pseudo-polar point groups. The point group symbols labeled by gray (black) color are non-centrosymmetric (centrosymmetric).

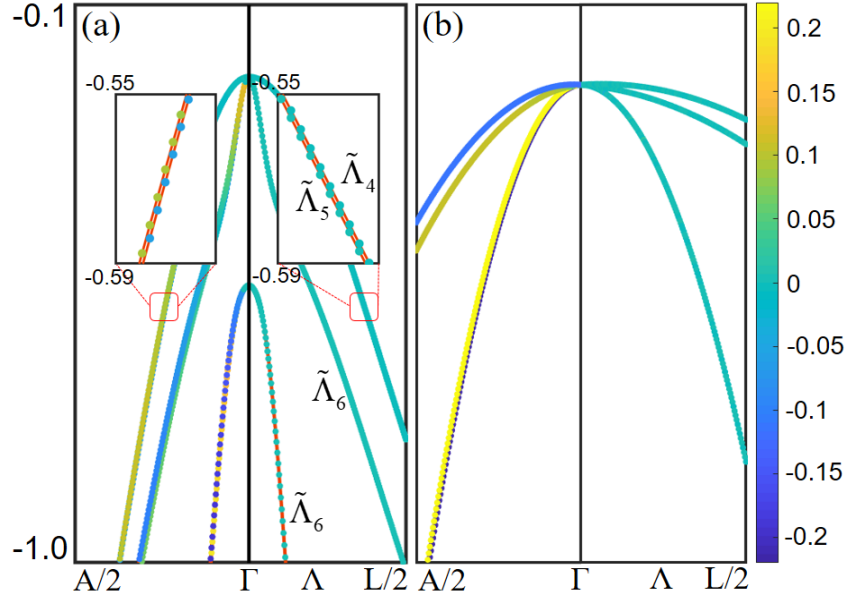

Figure 3. BSVSP in bulk GaAs. (a) Band structure and the *x* component of spin polarizations of bulk GaAs calculated by DFT. The *y* and *z* components have the same behavior. The coordinates of symmetry points A and L are (0.5,0.25,0) and (0.5,0.5,0.5), respectively. The irreducible representation $\widetilde{\Lambda}_4$ and $\widetilde{\Lambda}_5$ are one dimensional. The irreducible representation $\widetilde{\Lambda}_6$ is two dimensional. (b) Band structure and the *x* component of spin polarizations of bulk GaAs calculated by the k.p Hamiltonian.

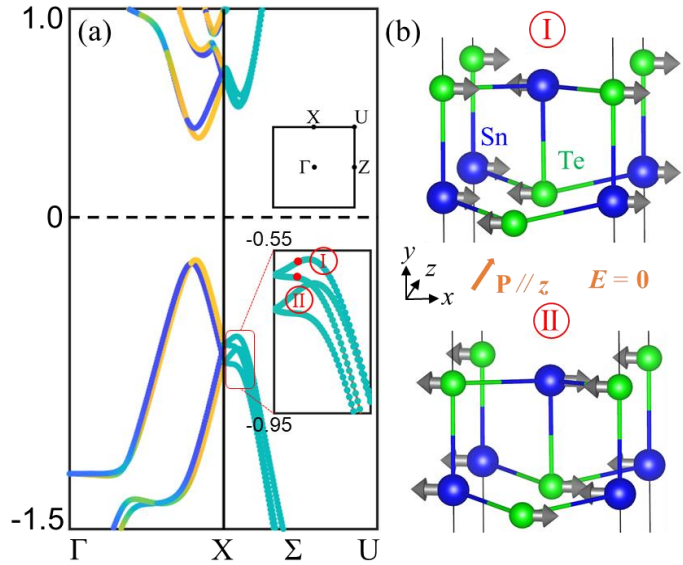

Figure 4. BSVSP in 2D SnTe thin film. (a) Band structure and the *y* components of spin polarizations of 2D SnTe thin film calculated by DFT. The *x* and *z* components all vanish and are not shown here. The color bar of the spin polarization is the same as that in Fig. 3. (b) Local spin polarizations of Sn and Te atoms for the Bloch states marked with the red points I and II in (a).

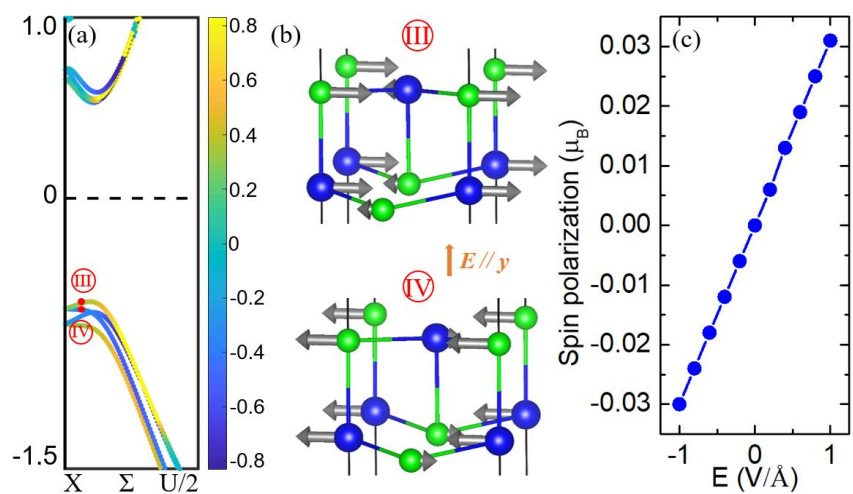

Figure 5. Linear magnetoelectric coupling in reciprocal space in 2D SnTe thin film. (a) The $x$ components of spin polarizations when a 0.1 V/Å external electric field parallel to $y$ axis (i.e., normal to the film) is added. The vanishing $y$ and $z$ components are not shown. (b) Local spin polarizations of Sn and Te atoms for the Bloch states marked with the red points III and IV in (a). (c) Dependence of $S_x$ on the external electric field for the Bloch state marked with the red point III.

Acknowledgments

Word at Fudan is supported by NSFC 11825403, the Special Funds for Major State Basic Research (Grant No. 2015CB921700), the Program for Professor of Special Appointment (Eastern Scholar), the Qing Nian Ba Jian Program, and the Fok Ying Tung Education Foundation. K.L. thanks Junsheng Feng for useful discussions.